\documentclass[12pt]{article}
\usepackage[latin9]{inputenc}

\usepackage{wrapfig}
\usepackage{units}
\usepackage{amsbsy}
\usepackage{amstext}
\usepackage{amsmath}
\usepackage{graphicx}
\usepackage{xcolor}

\usepackage{amssymb}

\makeatletter

\usepackage{times}

\newenvironment{sciabstract}{%
\begin{quote} \bf}{\end{quote}}

\title{Hidden Magnetic Texture in the Pseudogap Phase of High-Tc  $YBa_{2}Cu_{3}O_{6.6}$ }

\author{Dalila Bounoua$^{1\ast}$, Yvan Sidis$^{1}$, Toshinao Loew$^{2}$, Fr\'ed\'eric Bourdarot$^{3}$, Martin Boehm$^{3}$,\\  
Paul Steffens$^{3}$, Lucile Mangin-Thro$^{3}$,  Victor Bal\'edent$^{4}$,  and Philippe Bourges$^{1\ast}$\\
\\
\normalsize{$^{1}$ Universit\'e Paris-Saclay, CNRS-CEA, Laboratoire L\'eon Brillouin, 91191, Gif sur Yvette, France}\\ 
\normalsize{$^{2}$ Max Planck Institute for Solid State, Heisenbergstrasse 1, research 70569 Stuttgart, Germany}\\ 
\normalsize{$^{3}$ Institut Laue-Langevin, 71 avenue des Martyrs, Grenoble 38000, France}\\
\normalsize{$^{4}$ Universit\'e Paris-Saclay, Laboratoire de Physique des Solides,  Orsay 91405 CEDEX, France}\\
\\
\normalsize{$^\ast$To whom correspondence should be addressed, e-mail:}{dalila.bounoua@cea.fr, philippe.bourges@cea.fr}
\\
}

\makeatother

\begin{document}

 \baselineskip24pt


\maketitle

\begin{sciabstract}

Despite decades of intense researches,  the enigmatic pseudo-gap (PG) phase of superconducting cuprates remains an unsolved mystery.  In the last 15 years, various symmetry breakings in the PG state have been discovered, spanning an intra-unit cell (IUC) magnetism, preserving the lattice translational  (LT) symmetry but breaking time-reversal symmetry and parity, and an additional incipient charge density wave breaking the LT symmetry upon cooling. However, none of these states can (alone) account for the partial gapping of the Fermi surface.  Here we report a hidden LT-breaking magnetism uisng  polarized neutron diffraction. Our measurements reveal magnetic correlations, in two different underdoped $\rm YBa_{2}Cu_{3}O_{6.6}$ single crystals, that settle at the PG onset temperature with i) a planar propagation wave vector $(\pi,0) \equiv (0,\pi)$, yielding a doubling or quadrupling of the magnetic unit cell and ii) magnetic moments mainly pointing perpendicular to the  $CuO_{2}$ layers. The LT-breaking magnetism is at short range suggesting the formation of clusters of 5-6 unit cells. Together with the previously reported  IUC magnetism, it yields a hidden magnetic texture of  the $CuO_{2}$ unit cells hosting loop currents,  forming large supercells  which may be crucial for elucidating the PG puzzle. 

\end{sciabstract}



The phase diagram of high temperature cuprate superconductors is dominated  by the mysterious $PG$ phase overhanging the unconventional d-wave superconducting ($SC$) state  \cite{Keimer15,Proust19,Varma20}.  
A fundamental property of the $PG$ is a partially gaped electronic spectrum whose origin has been under hot debate and deep experimental and theoretical scrutinies since its discovery \cite{alloul1989}.  A wide set of experimental results report the onset of broken discrete Ising ($Z_{2}$) symmetries at the same characteristic temperature T* where the electronic  $PG$ opens \cite{bourges2021-online}.  The discrete broken symmetries are: lattice rotation ($C_{4}$) \cite{sato2017thermodynamic,lawler10,daou10}, interpreted in terms 
of an (Ising) nematic order,  inversion or parity ($P$) as shown by second harmonic generation  \cite{zhao2017global} measurements and time-reversal ($T$)  as reported by polarized neutron diffraction ($PND$)  \cite{fauque06,li2008unusual,Mook08,Baledent11,Bourges11,Mangin17,Jeong17}, circularly polarized angle resolved photo-emission spectroscopy \cite{Kaminski02} and 
muon spin spectroscopy \cite{zhang2018discovery}. The $T$ and $P$ symmetry  breakings are usually associated with a translationally invariant or Intra-Unit Cell magnetism ($q$=0 or IUC-magnetism), either carried by loop currents ($LCs$) \cite{Varma06,Agterberg15,Chatterjee17,Scheurer18,Sarkar19} or by magnetic multipoles \cite{Lovesey15,Lovesey15b,fechner2016quasistatic}. The detection of such objects is extremely challenging owing to the weakness of the corresponding signal, which is further
partly hidden by the structural response during a scattering experiment \cite{fauque06,li2008unusual,Mook08,Baledent11,Bourges11,Mangin17,Jeong17}. 
 However, to date, the reported $IUCs$ pattern preserves the $LT$ symmetry. A large Fermi surface is thus conserved and none of the $IUCs$ related broken discrete symmetries can induce the needed electronic gap \cite{Keimer15,Varma19}. Meanwhile, the Fermi electron pockets observed at high magnetic field and low temperature in the $PG$ phase \cite{Proust19,Varma20} must originate from a phase breaking $LT$ symmetry. 

\begin{figure}
\begin{centering}
\includegraphics[width=15cm]{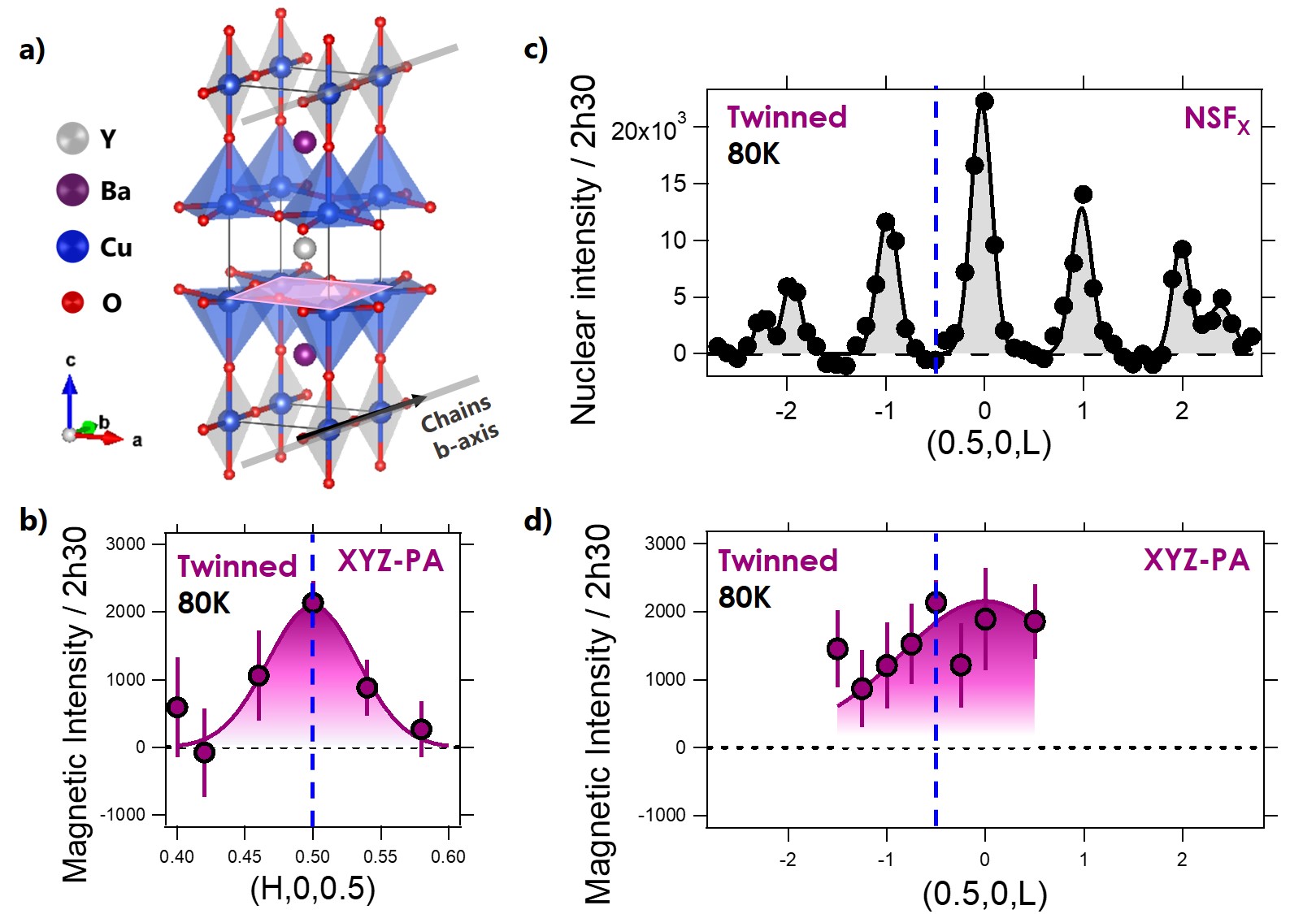} 
\par\end{centering}
\clearpage
\caption{\label{Fig1} \textbf{Novel commensurate magnetism in twinned $YBa_{2}Cu_{3}O_{6.6}$ (YBCO-t) }:
\textbf{(a)} Crystal structure of $YBa_{2}Cu_{3}O_{6+x}$ with the CuO chains running along the b-axis (gray shaded arrow) and the $CuO_{2}$ planes represented by the pink shading. 
\textbf{(b)}  H-scan across (0.5,0,0.5) showing magnetic scattering extracted from $XYZ-PA$. The magnetic scattering is fitted by a Gaussian line centered at (0.5,0,0.5).
\textbf{(c)} Background subtracted L-scan along the (0.5,0,L) direction in the non-spin flip ($NSF_{X}$) channel. The peaks in the ($NSF_{X}$) channel correspond to nuclear scattering from the Ortho-II oxygen chains superstructure.
\textbf{(d)} \textbf{Q} - dependence of the magnetic intensity along the (0.5,0,L) trajectory as extracted from full $XYZ-PA$ in YBCO-t.
Data in \textbf{(b-d)} were measured on  {$TAS-4F1$}  at {$80K$} with the sample aligned in the (1,0,0)/(0,0,1) scattering plane. Lines are fits to the data. Error bars represent one standard deviation. Raw data are given in supplementary information.
 }
\end{figure}

In the same region of the phase diagram, at somewhat lower temperature than T*, an incipient modulated charge density wave $(CDW)$ that competes with $SC$ and breaks the $LT$ and $C_{4}$ symmetries was  extensively reported \cite{Keimer15,Proust19}. The short range d-wave $CDW$ induces a lattice superstructure leading to a bi-axial response at incommensurate planar wave-vectors in reciprocal space $q_{CDW}$=(0,$\pm\epsilon$) and/or ($\pm\epsilon$,0), with  $\epsilon\sim0.3$ in reduced lattice units ($r.l.u$) in  
${\rm YBa_{2}Cu_{3}O_{6+x}}$ (YBCO). At zero magnetic field, $CDW$ modulations are quasi-$2D$ and weaken when entering the $SC$ state at $T\textless T_{c}$. Upon applying either an external magnetic field along the  $c$-axis  \cite{chang2016magnetic} or a uni-axial pressure along the $a$-axis \cite{kim2018uniaxial}, 3D CDW correlations develop with a uni-axial character along the  $b$-axis (direction of the underlying $CuO$ chain in YBCO, see  Fig. \ref{Fig1}.a). Although the incipient $CDW$ breaks $LT$ symmetry, its onset temperature $T_{CDW}$ remains well below 
$T^{*}$, deep into the $PG$ phase, and can thus not be alone at the origin of the opening of the electronic gap. This conclusion is further reinforced by recent Hall transport measurements \cite{badoux2016change}.   

To describe this manifold problem and reconcile the seemingly unrelated discrete symmetry breaking and occurrence of the electronic gap, new theoretical approaches invoking the concept of intertwined states were elaborated \cite{Keimer15}. Apart from $IUC-LCs$ magnetism and modulated $CDW$, several studies focused on the search for unconventional magnetic correlations involving alternative symmetry breaking charge current patterns in cuprates. For instance, starting from a collinear homogeneous antiferromagnetic ($AF$) Mott insulator (zero hole-doping, localized $S=\frac{1}{2}$ spins on $Cu$ atoms) with a magnetic response at the planar wave-vector $\bf q_{AF}$=(0.5,0.5), hole-doping was proposed to destabilize the AF state towards a spin-liquid state characterized by a staggered flux phase ($\pi$-flux) \cite{Hsu91}, also described as a $d$-wave charge density wave ($DDW$) state \cite{Chakravarty01}. This state exhibits staggered $LCs$, yielding an orbital magnetic response at $\bf q_{DDW}$=$\bf q_{AF}$.  Note, however, that the $Cu-$spin moments are locked within the $CuO_2$ layers (planar anisotropy) \cite{Regnault98}, whereas the $LCs$ orbital moments are expected to be perpendicular to the $CuO_2$ layers.  Despite several attempts, neutron diffraction measurements failed to prove the existence of such a phase \cite{Bourges11,Stock02}. 

Here, we report the discovery of magnetic correlations at commensurate planar \textbf{Q}-wave-vectors $(q,0)\equiv(0,q)$ with $q=1/2$ that appear to be tightly bound to the $PG$ physics in the underdoped  $\rm YBa_{2}Cu_{3}O_{6.6}$ material.
We performed elastic PND experiments on three Triple Axis Spectrometers (TAS): 4F1 at Orph\'ee reactor-Saclay, IN22 and Thales at Institut Laue-Langevin-Grenoble, operating at different wavelenghts and using different polarization setups (see methods). The TAS were equipped with longitudinal XYZ polarization analysis ($XYZ-PA$), a powerful technique to selectively probe and disentangle the magnetic response from the nuclear one with no assumptions about on the background. The samples were studied either in (1,0,0)/(0,0,1) or  (1,0,0)/(0,1,0) scattering planes such as wave-vectors of the form $(H,0,L)$ or $(H,K,0)$ were accessible. In the following, the wave-vectors are indexed in reduced lattice units ($r.l.u$), in units of $(\frac{2\pi}{a}, \frac{2\pi}{b},\frac{2\pi}{c})$ where $a$, $b$ and $c$ stand for the lattice parameters of the sample (See Methods for details). 
We carried out our investigations in two different samples: a twinned YBCO-t sample, same as \cite{fauque06}, 
and a detwinned sample YBCO-d, same as \cite{fauque06,Mangin17},  with close compositions (nominal $x$=0.6 and hole-doping $p\sim 0.12$) and critical temperatures ($T_c$=61 and 63 K) but with different oxygen ordering of the CuO chains. 

%
Using  $XYZ-PA$, we uncovered a magnetic peak centered at \textbf{Q}=(0.5,0,0.5) in YBCO-t, as shown by the H-scan along (H,0,0.5) (Fig. \ref{Fig1}.b). The magnetic signal can be described by a Gaussian profile, broader than the instrumental resolution.  After deconvolution, its intrinsic linewidth  (half width at half maximum) is $\Delta_H$=$0.03\pm0.01$ r.l.u. This corresponds to a finite correlation length along the $a$-axis of  $\xi_{a}=\frac{a}{2\pi \Delta}$= $\sim20\pm6\text{\AA}$ , corresponding to $\sim 5$ planar unit-cells. A rocking scan across (0.5,0,0.5) allowed us to confirm the intrinsic origin of the magnetic signal, excluding extrinsic scattering due to powder lines from  a parasitic magnetic phase (see supplementary information). 

The $L$-scan of Fig. \ref{Fig1}.c  was performed in the $NSF_X$ channel and shows a nuclear scattering at \textbf{Q}-positions of the form (0.5,0,L) with $L$ integer. This peculiar structure factor is due to the characteristic $Ortho-II$ structure of the $CuO$ chains, running parallel to the $b$-axis as represented in Fig. \ref{Fig1}.a  \cite{andersen1999superstructure}. 
The same scan along  $(0.5,0,L)$ direction, obtained using  $XYZ -PA$, shows a broad magnetic scattering,  at odds with the $Ortho-II$ oxygen ordering in the $CuO$ chains (Fig. \ref{Fig1}.d and supplementary information). This rules out the chains as the host of the new-found magnetic signal rather pointing to the $CuO_{2}$ planes (see  Fig. \ref{Fig1}.a) as the origin of the magnetic response, breaking the $LT$  invariance ($q=1/2$). The diffuse magnetic signal along the $(0.5,0,L)$ trajectory
of Fig. \ref{Fig1}.d further underlines the absence of magnetic correlations along the $c$-axis.

%

\begin{figure}
\begin{centering}
\includegraphics[width=15cm]{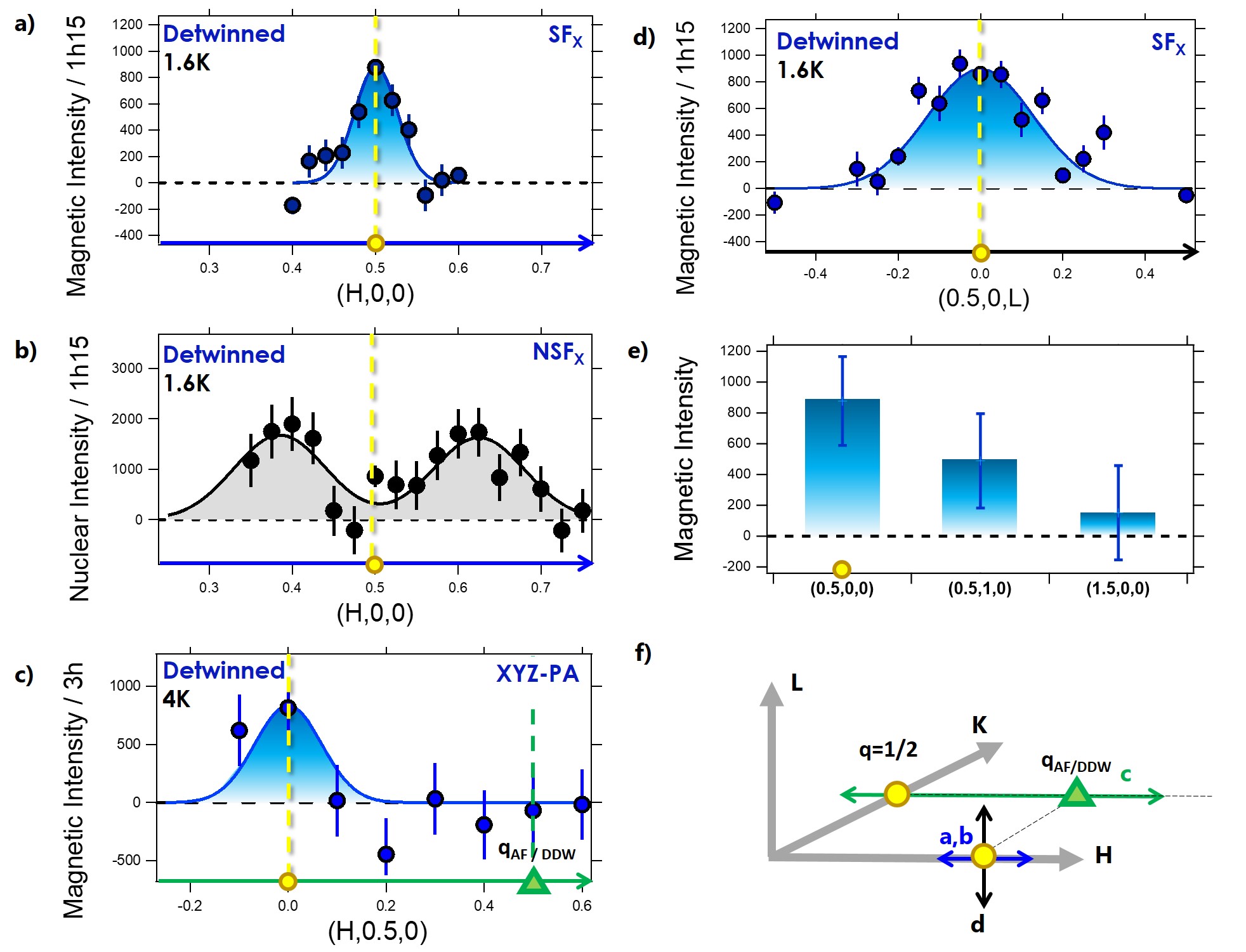} 
\par\end{centering}
\clearpage
\caption{\label{Fig2} \textbf{Biaxial planar short range magnetism in twin-free $YBa_{2}Cu_{3}O_{6.6}$ (YBCO-d)}:
\textbf{(a)}Background subtracted H-scan across (0.5,0,0) in the Spin Flip ($SF_{X}$) channel. The magnetic intensity appears as a Gaussian signal centered at (0.5,0,0).
\textbf{(b)} Background subtracted H-scan along (H,0,0) in the Non Spin Flip ($NSF_{X}$) channel showing the background subtracted nuclear scattering from the (Ortho-VIII) type chains superstructure. 
\textbf{(c)}  H-scan across the (H,0.5,0) showing a magnetic singal centered at $H=0$. 
\textbf{(d)} Background subtracted L-scan across (0.5,0,0) in the spin-flip ($SF_{X}$) channel. The magnetic intensity appears as a Gaussian signal centered at (0.5,0,0). 
{\textbf{(e)} Magnetic intensity resulting from $XYZ-PA$ at different \textbf{Q}-points in reciprocal space : (0.5,0,0), (0.5,1,0) and (1.5,0,0)}. 
\textbf{(f)} 3D representation of the reciprocal space showing the momentum scans performed in \textbf{(a-d)}. 
Data in \textbf{(a-b,d)} were measured on {$IN22$}  at {$1.6K$} with the sample aligned in the (1,0,0)/(0,0,1) scattering plane. {Data in \textbf{(a-b,d)} were measured on {$IN22$}  at {$1.6K$} with the sample aligned in the (1,0,0)/(0,1,0) scattering plane.}.  Data in \textbf{(c)} were measured on {$4F1$} at {$4K$}  with the sample aligned in the (1,0,0)/(0,1,0) scattering plane. Lines are fits to the data. Error bars represent one standard deviation. Raw data are given in supplementary information. 
}
\end{figure}

To ascertain the existence of the $q=1/2$ magnetism and unravel scattering from $CuO$ chains and $CuO_{2}$ planes, we carried further $PND$ investigations in YBCO-d along a few directions in momentum space shown in Fig.\ref{Fig2}.f.
Fig. \ref{Fig2}.a shows a H-scan in the $SF_X$ channel revealing the occurrence of a  magnetic peak, centered at (0.5,0,0) $\equiv (\pi,0)$. As in YBCO-t, the signal is at short range with $\Delta_H$= 0.025 $\pm$ 0.01 r.l.u. given by the fit with a Gaussian profile. This corresponds to $\xi_{a}=\frac{a}{2\pi \Delta}$= $\sim24\pm4\text{\AA}$, which represents correlations over $\sim6$ unit cells, consistent with the value deduced in the YBCO-t sample.  
The same scan in the $NSF_X$ channel Fig. \ref{Fig2}.b reveals two nuclear peaks at $H=0.385$ and $0.625$, inherent to the $Ortho-VIII$ oxygen ordering of the $CuO$ chains 
 \cite{ andersen1999superstructure} in that YBCO-d sample  \cite{hinkov2004two}, which leads to a nuclear contribution at $q_{Ch}=(H\pm0.125,0,0)$ where no magnetic signal occurs (Fig. \ref{Fig2}.a). The absence of magnetic scattering at the characteristic $q_{Ch}$ positions confirms the $CuO_{2}$ planes as the origin of the magnetic response at  $H=0.5$. 

We further performed a survey of the momentum dependence of the magnetic signal in the detwinned YBCO-d sample, where $a$ and $b$ directions are clearly identified, to determine its planar structure factor. Fig. \ref{Fig2}.c shows the resulting magnetic intensity along the $(H,0.5,0)$ trajectory, as given by $XYZ-PA$. 
Interestingly, the $H$-scan reveals as well a magnetic peak centered at $(0,0.5,0) \equiv (0,\pi)$  while no magnetic intensity is seen at $\bf q_{AF/DDW}$=$(0.5,0.5,0)\equiv (\pi,\pi)$. The associated magnetic pattern is then found along both $CuO$-bond directions with actually a very similar intensity. That indicates a short range uni-axial (with domains) or bi-axial magnetism leading to a local doubling or quadrupling (2x2) of the unit cell, never reported previously. Additional measurements with the sample aligned within $[a,c]$ plane allowed us to elucidate the $L$-dependence of the out-of-plane magnetic correlations. Fig. \ref{Fig2}.d shows an L-scan across $(0.5,0,0)$ in the $SF_{X}$ channel highlighting a broad magnetic peak, centered at $(0.5,0,0)$. The extracted correlation length  (from a fit by a Gaussian function) $\xi_{c} \simeq 13\pm1\text{\AA}$ is very short and does not exceed  $\sim 1$ unit-cell in the $c$ direction  ($\Delta_L$= $0.14\pm0.01$ r.l.u). 

We further investigated the structure factor of the (2x2) magnetism through $XYZ-PA$ measurements at  additional \textbf{Q}-points. The results are reported in Fig.\ref{Fig2}.e and show the occurence of magnetic intensity at wavevectors of the form (0.5,1,0) as well, whereas the limited statistics at (1.5,0,0) prevents from drawing a clear conclusion about the existence of a magnetic response and requires further investigations.


To determine the onset temperature of the short range $q=1/2$ magnetism, we measured the temperature dependence of the magnetic signal at  $(0.5,0,0)$ in YBCO-d, both in the $SF_{X}$ channel and using $XYZ-PA$. Fig. \ref{Fig3}.a shows the total scattered magnetic intensity that splits into a leading out-of-plane magnetic component $I_{c}$  displaying an order parameter-like temperature dependence in Fig.\ref{Fig3}.b and a subsidiary in-plane magnetic component, $I_{b}$, which at variance, remains almost constant as a function of temperature Fig.\ref{Fig3}.c. Our measurements show that the onset temperature for the dominating $I_{c}$  magnetic component is $T^{*}\sim235K$. The temperature dependence of the magnetic signal at $(0,0.5,0)$ as deduced from $XYZ-PA$ analysis reproduces the same Ising-like orientation of the magnetic moment, with a similar amplitude as at  $(0.5,0,0)$, with a major $I_{c}$, out-of-plane component arising at $T^{*}\sim235K$ and following the same order parameter-like temperature dependence (see supplementary information). The H-scans across $(0.5,0,0)$ on Figs. \ref{Fig3}.{d-e} show the drop of the magnetic signal from $10$ to $300K$, respectively. The data correspond to measurements in the $SF_{X,Y,Z}$ channels. Note in  Fig. \ref{Fig3}.{d} that  the magnetic signal in the $X$ channel corresponds to 6 times the statistical noise related to the background. Further, since there is no atomic Bragg peak to subtract, this experiment is much easier than the earlier experiments on $IUC$-magnetism\cite{fauque06,Bourges11}. However, the nuclear scattering from the chains oxygen atoms is strong enough to hide the weaker magnetic scattering  in an unpolarized neutron experiment. 

Importantly, the temperature dependence of the out-of-plane magnetic response, $I_{c}$, matches the one reported for the $IUC$-magnetism (see supplementary information):  $I_{c}$ sets-in at  $T\simeq T^{*}$, the PG onset temperature as reported from resistivity measurements \cite{ito1993systematic}, and of $T\sim T_{mag}$, the onset temperature for the $q=0$ order \cite{fauque06,Bourges11}, suggesting a common origin of both  IUC order and short range $q=1/2$ magnetism reported here. On general grounds, because it breaks $LT$, the  $q=1/2$ magnetism could be an alternative candidate for the opening of the $PG$ state, potentially solving a long-standing question.  Along the same line of thought, it could as well play a significant role in the electron Fermi pockets formation \cite{Proust19} yielding a Fermi surface reconstruction known to be exclusively occurring within the $PG$ state. 

%
%
Our comprehensive set of PND data gives a coherent picture for the  $q=1/2$  magnetism occuring at commensurate $(\pi,0)\equiv (0,\pi)$  in high-$T_{c}$ superconducting $YBa_{2}Cu_{3}O_{6.6}$. First, the location of the observed hidden magnetism is clearly distinct from $q_{AF}\equiv(\pi,\pi)$. This rules out various reported magnetic patterns, namely the $AF$  spin order, $\pi$-flux phases-like patterns or  $DDW$ \cite{Hsu91,Chakravarty01}, all phases located at the 
commensurate $(\pi,\pi)$ position, or Spin Density Wave (SDW) located at incommensurate position  $q_{SDW}$=$q_{AF}\pm(\delta,0)$ or $(0,\delta)$ with $\delta\sim 0.1$ revealed by spin fluctuations induced by zinc impurity \cite{Suchaneck10}. 
Second, the occurence of both propagation vectors ${\bf q_{1}}=(0.5,0,0)$ and ${\bf q_{2}}=(0,0.5,0)$ suggests a biaxial ordering with a double-$\bf q$ antiferromagnetic structure, corresponding to a quadrupling (2x2) of the unit cell. However, uniaxial ordering, doubling the unit cell along each direction of the CuO bonds, can as well account for the observed patterns, assuming equi-populated magnetic domains along each direction. This can happen even in the detwinned sample if the magnetic domains do not depend on the underlying orthorhombic structure. 

We considered several models that can reproduce the scattering selection rules for the $q=1/2$ magnetism. Fig. \ref{Fig4} shows three different magnetic patterns corresponding to $2\times2$ larger unit cells and involving either: spin or orbital magnetic moments at one $Cu$-site over 2  (Fig. \ref{Fig4}.a) with an antiferromagnetic coupling along the unit cell diagonal, a loop currents model  (Fig. \ref{Fig4}.b) corresponding to the smallest possible domain (P=1) of $LC$s supercell recently proposed to break the lattice translation symmetry  \cite{Varma19} and where the anapole undergoes a  $90^{\circ}$ rotation at each adjacent unit cell \cite{bourges2021-online}, or a staggered loop currents phase with currents running along the diagonals of the $\rm CuO_2$ planes  (Fig. \ref{Fig4}.c). All these patterns nicely agree with the measured structure factor (see supplementary information). However, the magnetic moment predominantly observed perpendicular to the $CuO_2$ layers does not support an interpretation in terms of $Cu$ spins that are locked within the $CuO_2$ planes owing to their strong $XY$ anisotropy \cite{Regnault98}. It instead favors loop current patterns as the ones shown in Figs. \ref{Fig4}.b-c where orbital moments have to be perpendicular to the planes where currents are confined \cite{Varma06}.

Following all these observations, one can build real space pictures based on loop currents to account for both IUC order and hidden $q=1/2$ magnetism. As originally proposed in \cite{Simon02,Shekhter09}, LCs are conveniently  characterized by four degenerate states that can be represented by anapole moments pointing along the four planar unit cell diagonals  as displayed in  Fig. \ref{Fig5}.a where each color corresponds to a given anapole orientation. Using these four basic states, one can build an anapole-vortex-like phase represented in Fig. \ref{Fig5}.b that also describes the  $q=1/2$  magnetism. It is similar to Fig. \ref{Fig4}.b but now with anapoles located on the $Cu$-site \cite{Varma19}.  The pattern is both chiral and $P$-breaking, and therefore of $mm2$ symmetry
as it has been recently reported from photogalvanic experiments \cite{Lim20}. The amplitude of the total magnetic scattering has been estimated in absolute units after calibration by the intensity of the reference (1,0,0) nuclear Bragg peak. This magnetic intensity is weak and represents only $\sim 0.3\pm 0.1$ mbarn once integrated in momentum space, which is about 10-15 times lower than the one reported for the long range ordered ($q=0$) $IUC$-magnetism at the (1,0,0) position \cite{fauque06,Mook08,Bourges11,Mangin17}. This explains why the magnetic signal remained hidden in previous experiments.  To give an order of magnitude, this would correspond to an estimate of the magnetic moment per loop current triangle of $\sim$ 0.02 $\mu_B$ considering the pattern of Fig.\ref{Fig5}.b   under the assumption of a homogeneous distribution of moments where all unit cells contribute to the magnetic signal. However, as the $q=1/2$ magnetism is at short range,  this hypothesis  is not consistent with the data, leading instead  to a picture that one unit-cell over four contributes to the $q=1/2$ magnetism  (see below), leading therefore to a much larger moment of $\gtrsim$ 0.08  $\mu_B$ per unit cell.

A natural question is then what can be the interplay between both LCs phases with distinct propagation wave-vectors, $q=0$ and $q=1/2$ ? First, the short range nature of the observed correlations implies $2\times 2$ $LCs$ islands of only about $\sim$ 20-25 \AA. This therefore precludes a uniform distribution of $2\times 2$ $LCs$ superposed to LCs that respect LT.  Inhomogeneous pictures in real space should be considered instead, revealing a hidden anapolar or magnetic texture. How can a unifom LC order coexist with a modulated one at short range ? This  can be accounted for by locating the short range modulated phase at the domains walls between the long range uniform ($q=0$) domains.  Along the same line of thought and by analogy with the framework of electronic liquid crystals of coexisting smectic modulations and intra-unit-cell nematicity \cite{mesaros2011}, one can speculate about the coexistence of a "smectic" short range ($q=1/2$) $2\times 2$ $LCs$ magnetism with a "nematic" ($q=0$) longer range magnetism (ferro-anapolar). 


According to this picture,  the observed amplitudes of the $q=1/2$ magnetism (located at the boundaries separated larger $q=0$ domains) would determine the volume fraction of both magnetic patterns within the crystal.  In contrast to the uniform picture, the 10-15 times weaker intensity of the q=1/2 magnetism instead implies that one site over four belongs to the $2\times 2$ $LCs$ clusters assuming the same anapole amplitude  (and so, the same magnetic moments amplitude)  in all unit cells. Real space pictures can be built following these requirements.  As an example,  Fig. \ref{Fig5}.c shows a pattern of anapoles over 20x20 unit cells that represents such a topological arrangement of anapoles (and consequently related loop currents and orbital magnetic moments). Its calculated structure factor matches the observations of the short range magnetism at both (0.5,0,0) and (0,0.5,0) and longer range IUC magnetism. This picture can be extended  to a larger number of unit cells (for instance with more than one $2\times 2$ $LCs$ cluster) with only two requirements: i) the clusters should be isolated and ii) should represent in total a quarter of the unit cells. The longer range uniform LCs domains are forming the large supercells (of $2P\times2P$ size) introduced by C.M. Varma \cite{Varma19} to account for the Fermi arcs in the PG state. Note however that the P=1 reported here cannot explain the observed Fermi arcs or magneto-oscillations. LC supercells are predominantly characterized in momentum space by a set of satellites magnetic peaks located at $\pm{1\over{2P}}$ from the atomic Bragg peak positions. As far as $P \gtrsim 10$, they would appear at $q=0$ due to the limited momentum instrumental resolution of elastic PND experiments \cite{bourges2021-online}. However, more experiments are necessary to clarify this point.  Meanwhile,  as far as 1 to 4 site occupation rule holds, 1D arrangements of anapoles either horizontal or vertical as shown in Fig. \ref{Fig5}.d-e could as well describe the data, with however an additional constraint of equi-populated domains along both directions. Note that the magnetic texture, through the domains size  $P$, can vary noticeably within the crystal leading to a distribution of satellite peaks near the Bragg peaks with similar  short range $q=1/2$  magnetism. One can therefore envisage a vast range of possible hidden magnetic textures opened by the observation of short range  $q=1/2$ magnetism.

 About the interplay of  $q=0$ and $q=1/2$  magnetism, two interesting points should be emphasized. First, the data for  $q=1/2$  magnetism shows a larger moment along the $c$-axis whereas the moments of  ${\bf q}=0$ are tilted with respect to the $c$-axis \cite{Bourges11,Tang18}. It should be stressed that it can be consistent as the planar contribution of the magnetic structure factor
 for both signals depends largely on the specific modelling of the planar component whose origin is still under discussion \cite{bourges2021-online}. Next, the $q=0$  magnetism shows a planar anisotropy of the neutron intensity along both  in-plane directions, suggesting a specific arrangement of anapoles within the bilayer \cite{mangin2017}. Instead, the short range  ${\bf q} =1/2$ domains show an isotropic signal. Here again, the LCs correlations within the bilayer can be different for both signals because the 2x2 $LCs$  patterns are mixing the 4 anapolar directions. Clearly, more experimental data are necessary to settle these questions.

Finally, it is worth recalling that magnetic local probes experiments (Nuclear Magnetic Resonance and muon Spin Rotation) do not observe the expected static local magnetic fields of the magnetism reported by elastic PND measurements \cite{bourges2021-online,Wu15}.  Instead, a fluctuating magnetic response in the PG state has been reported by muon spectroscopy\cite{zhang2018discovery}. This suggests a magnetic texture slowly fluctuating at a $\sim 10 ns$ time scale, encompassing the ${\bf q}=0$ IUC response and the short range ${\bf q} =1/2$ magnetism. Both appear static in PND owing to the instrumental energy resolution of  $\sim$ 0.1 meV. Our discovery  of a novel ${\bf q} =1/2$  magnetic response, that in addition to the  previously reported ${\bf q}=0$ signal,  belongs to the magnetic structure factor of a complex  hidden magnetic texture (with large supercells of length scale $2P$ and magnetic moment of $\sim 0.1 \mu_B$ in each unit cell). We hope it would motivate further experimental investigations in other cuprate families and other hole-doping levels as well as theoretical investigations to understand the role of the hidden magnetic texture in the phase diagram of High-$T_{c}$ superconducting cuprates.

\section*{Methods} 

 \subsection*{Samples} 
The PND study was performed on two different samples of same nominal composition $\rm YBa_{2}Cu_{3}O_{6.6}$ although they exhibit different oxygen ordering. 
The twinned sample ($T_{c}$=61K and hole-doping p=0.107) is the same as the one used in \cite{fauque06} with an Ortho-II oxygen ordering. It has a mass of $\approx$10g and a mosaic spread of 1.5\textdegree, determined by performing rocking curves about nuclear Bragg peaks. 
The twin-free sample has been obtained by methods described in \cite{hinkov2004two}  ($T_{c}$=64K and hole-doping p=0.115). It consists of co-mounted single crystal plates of about (20 x 20 $mm^2$) and exhibits an Ortho-VIII oxygen ordering. The total sample mass is $\approx$2g and the measured mosaic spread is 2\textdegree. 
We used pseudo-tetragonal notations for the twinned sample, with  $a=b=3.85 \text{\AA}$ and $c=11.75 \text{\AA}$. 
The lattice parameters of the twin-free sample are  $a=3.88 \text{\AA}$, 
 $b=3.82 \text{\AA}$ and $c=11.75 \text{\AA}$. 

\subsection*{Polarized neutron diffraction}

The PND experiments were carried out on three instruments: the triple axis spectrometer (TAS)
4F1 (Orph\'ee reactor, Saclay) and the IN22 and Thales TASs (Institut Laue 
Langevin, Grenoble), which are described in more details in
 the Supplementary Materials \textbf{1}.  These instruments are equipped with distinct neutron
polarization set-ups and were operating with three distinct neutron
wavelengths, to guarantee the reproducibility of the measurements.
We used incident neutron wave-vectors
of ${\bf k_{i}}=2.57 \text{\AA}^{-1}$ on 4F1, ${\bf k_{i}}=2.662 \text{\AA}^{-1}$
on IN22, and ${\bf k_{i}}=1.5 \text{\AA}^{-1}$ on Thales. As detailed in the figures caption, the samples were either aligned in 
the (1,0,0)/(0,0,1)  or (1,0,0)/(0,1,0) scattering planes, so that wave-vectors \textbf{Q
}of the form $(H,0,L)$ or  $(H,K,0)$ were accessible, respectively. 
H, K or L-scans were performed across positions of the form (0.5,0,0) or (0,0.5,0) in r.l.u. 
Both $SF$ and $NSF$ scans were done in order to crosscheck the absence of nuclear scattering at magnetic
positions. 
The longitudinal $XYZ-PA$ was performed using:\\
-  MuPad spherical polarization analysis device with zero field at sample chamber on 4F1, for the twin-free  $YBa_{2}Cu_{3}O_{6.6}$ sample.  The incoming and outgoing beam polarizations are realized using Bender supermirrors. \\
- CRYOPAD spherical polarization analysis device with zero field sample 
chamber on IN22 and Thales for the twin-free  $YBa_{2}Cu_{3}O_{6.6}$ sample.  The incoming oand outgoing beam polarizations are realized using Heusler crystals. \\

\section*{Data availability}
The data obtained on IN22 at ILL are available at https://doi.org/10.5291/ILL-DATA.CRG-2776.
The data obtained on Thales at ILL are available at https://doi.org/10.5291/ILL-DATA.4-02-600.
The rest of the data that support the findings of this study is available 
from the corresponding authors upon request.
\bibliographystyle{nature}
\bibliography{papier}
\section*{Acknowledgments}

We thank C. P\'epin and Y. Zaanen for stimulating discussions. We thank  C.M Varma for bringing to us the illuminating perspectives of loop currents supercells.  We acknowledge supports from the project NirvAna (contract ANR-14-OHRI-0010)  of the French Agence Nationale de la Recherche (ANR) and from the GenLoop 
project  of the LabEX PALM (contract ANR-10-LABX-0039-PALM).

\section*{Author Information}
\subsection*{Contributions}
Y.S. and P.B. conceived and supervised the project; D.B., Y.S. and P.B. performed the experiments at LLB Saclay; D.B., V.B. and L.M.-T. performed the experiments at ILL Grenoble; F.B., M.B. and P.S. were the intrument local contacts and ILL. D.B. and P.B. analyzed the neutron data; T.L synthesized the single crystal samples; D.B., Y.S. and P.B. wrote the manuscript with further contributions from all authors. All authors contributed to this work, read the manuscript and agree to its contents.

Corresponding authors: Dalila Bounoua or Philippe Bourges.
\section*{Ethics declarations}
\subsection*{Competing interests}
The authors declare no competing interests.


%

\begin{center}
\begin{figure}
\begin{centering}
\includegraphics[width=15cm]{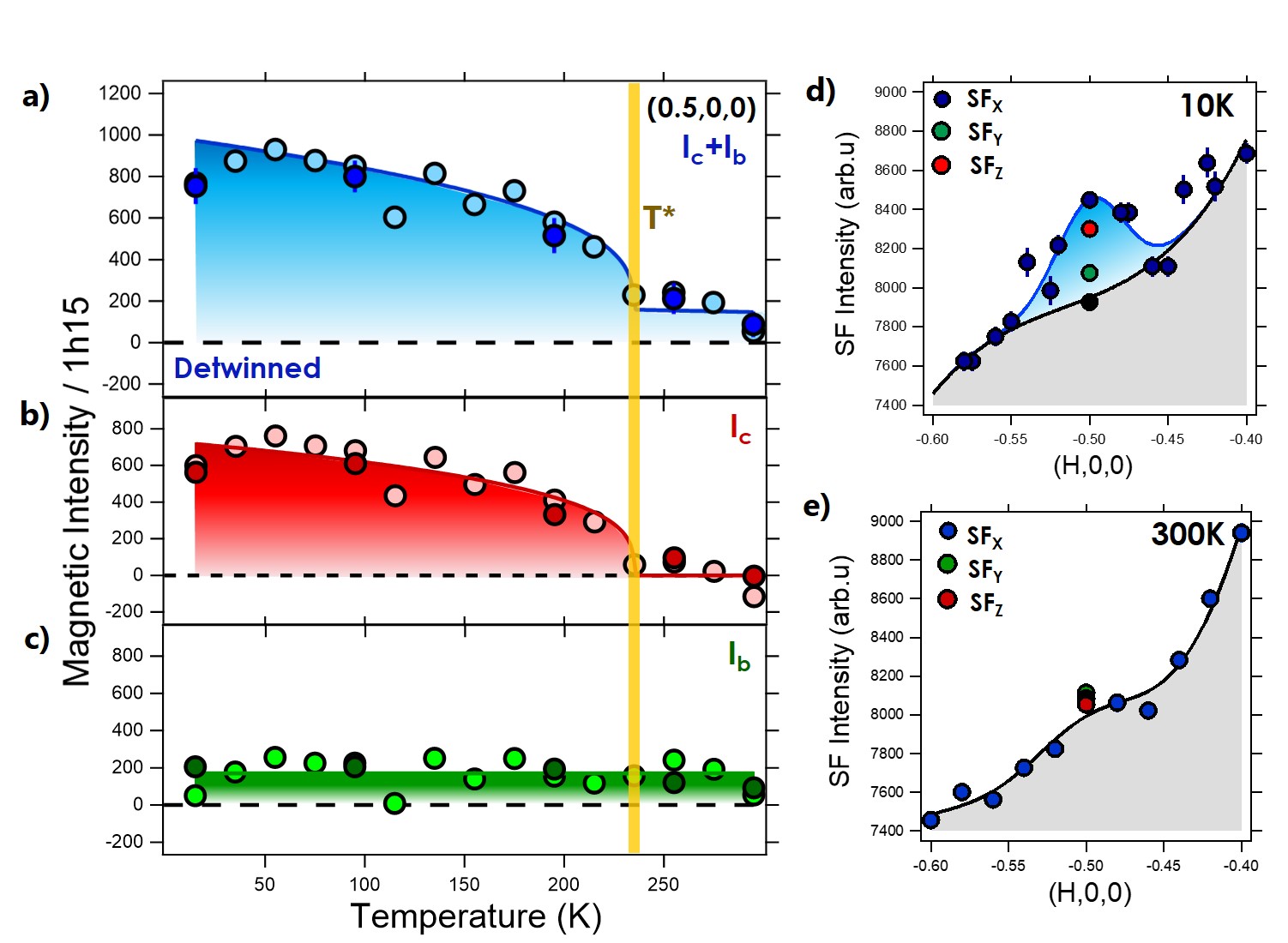} 
\par\end{centering}
\caption{\label{Fig3} \textbf{Temperature dependence of the biaxial magnetism in detwinned $YBa_{2}Cu_{3}O_{6.6}$ (YBCO-d)}:
 {\textbf{(a)} Temperature dependence of the magnetic intensity measured at  (0.5,0,0) as extracted from $XYZ-PA$ (dark blue circles) and background subtracted $SF_{X}$ data (light blue circles). Temperature dependence of \textbf{(b)} $I_{c}$, the out-of-plane magnetic response (dark red circles) and  \textbf{(c)} the in-plane magnetic scattering  $I_{b}$ (dark green circles) as extracted from $XYZ-PA$. Light red circles in \textbf{(b)} correspond to $I_{c}$ as extracted from the $SF_{X}$ data in \textbf{(a)} subtracted from the fit to the in-plane component in \textbf{(b)}.  Light green circles in \textbf{(c)} correspond to $I_{b}$ as extracted from the $SF_{X}$ data in \textbf{(a)} subtracted from the fit to the in-plane component in \textbf{(b)}. 
The blue symbols in panel \textbf{(a)} correspond to the sum $I_{c}+I_{b}$. H-scans across (0.5,0,0) measured in the Spin Flip ($SF_{X,Y,Z}$) channels at \textbf{(d)} 10K and \textbf{(e)} 300K. The magnetic intensity appears as a Gaussian signal centered at 10K and drops at 300K in YBCO-d}.
Data in \textbf{(a-e)} were measured on {$Thales$} with the sample aligned in the (1,0,0)/(0,0,1) scattering plane. Error bars (sometimes smaller than the points size) represent one standard deviation. Raw data  of \textbf{(a,b)} are given in supplementary information. 
}
\end{figure}%
\par\end{center}

\begin{center}
\begin{figure}
\includegraphics[width=14cm]{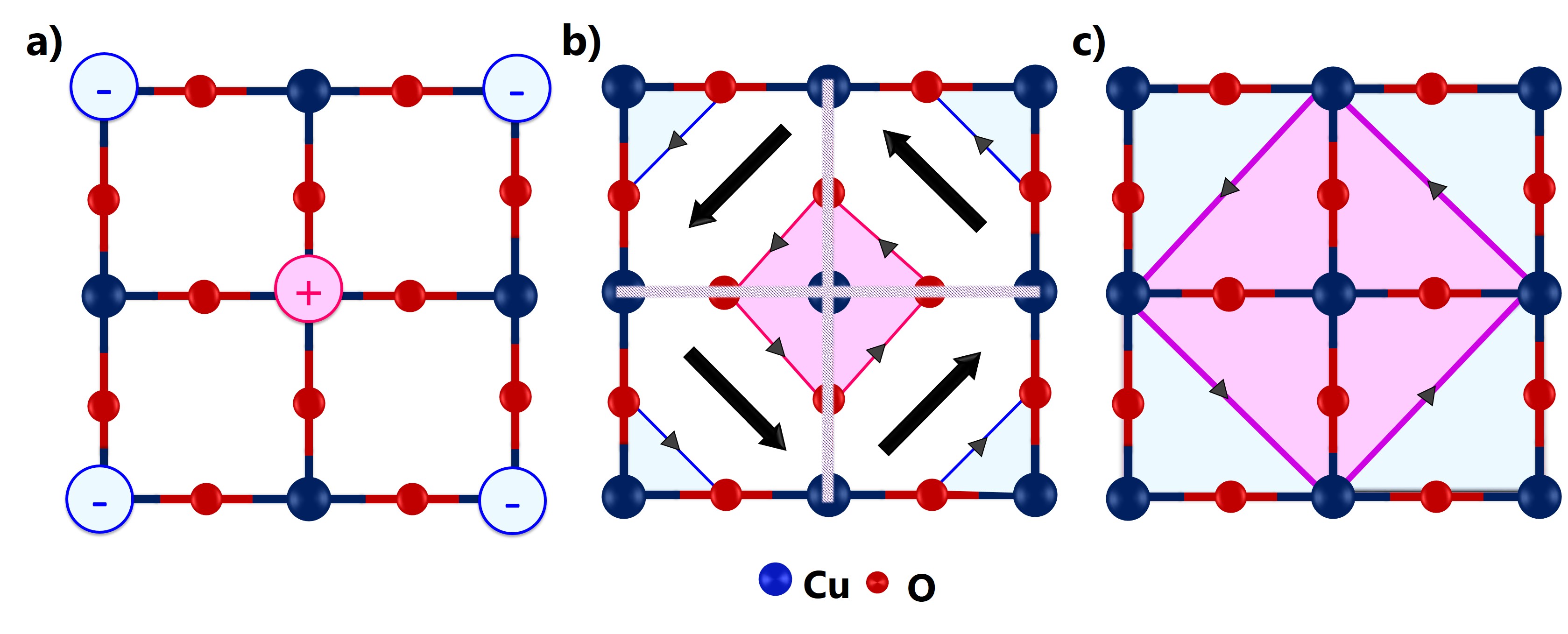} 
\caption{\label{Fig4} \textbf{Possible models for the biaxial (2ax2a where a is the cell parameter) planar magnetism in $YBa_{2}Cu_{3}O_{6.6}$}: 
\textbf{(a)} Magnetic moments at one over two Cu-site. \textbf{(b)} Loop currents pattern turning clockwise (in blue) and anti-clockwise (in pink), corresponding to P=1 in \cite{Varma19}. Each cell (size a x a) carries an anapole moment (black arrow). The pattern comprises 4 loop currents states where the anapole undergoes a $90^{\circ}$ rotation between adjacent cells.   \textbf{(c)}  Loop currents pattern with  $a$$\sqrt{2}$x$a$$\sqrt{2}$  loop size, consisting in a $45^{\circ}$ rotation of the {$DDW$} model with currents running between Cu-sites. All models in \textbf{(a-c)} reproduce the experimentally measured structure factor (see supplementary file). 
}
\end{figure}%
\par\end{center}

\begin{center}
\begin{figure}
\includegraphics[width=14cm]{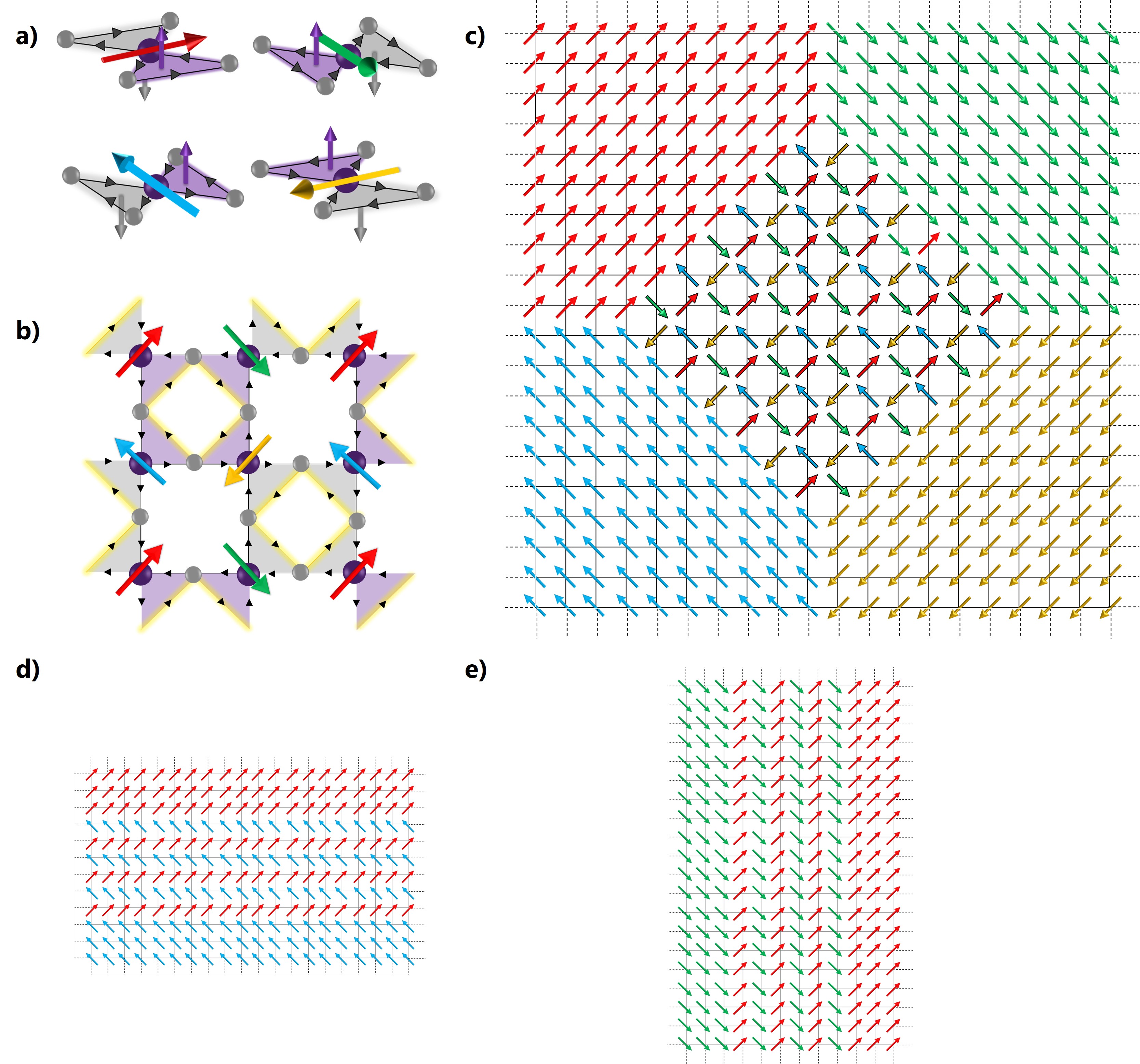} 
\caption{\label{Fig5} \textbf{Hidden magnetic texture from modelization of   $q=1/2$ short range and IUC magnetism observed in PND.} \textbf{(a)}  Four possible degenerate ground states of loop currents \cite{Simon02,Varma06,Shekhter09}.  The grey and purple arrows represent magnetic moments along the {\bf c} axis whereas the four other arrows represent anapoles centered at the Cu-site  of each of the four states. \textbf{(b)}  2x2 loop currents  pattern that can account for the $q=1/2$ magnetism. The currents are circulating clockwise (in grey) and anti-clockwise (in purple).  The four states are represented by  anapoles undergoing  a $90^{\circ}$ rotation between adjacent domains. \textbf{(c)} Example of 2D magnetic texture with 20x20 unit cells paved by anapoles (LCs states). The central cluster with  2x2 LC patterns describe the   $q=1/2$ short range magnetism whereas the IUC magnetic signal arises from the larger color domains. \textbf{(d,e)} Similar LCs construction for horizontal and vertical 1D domains. In panels \textbf{(c-e)}, only anapoles are represented having the four possible orientations. 
}
\end{figure}%
\par\end{center}

\end{document}